\documentclass[apjl]{emulateapj}
\def\mt{ }
\def\mtav{  }

\def\sas{ }
\def\sass{}

\usepackage{ifpdf}

\usepackage{makeidx}

\usepackage{amssymb}

\usepackage{amsmath}

\usepackage{eurosym}

\usepackage{amsfonts}

\usepackage{rotating}

\usepackage{mathtools}

\DeclarePairedDelimiter{\abs}{\lvert}{\rvert}
\usepackage{amssymb}

\def\cyg{Cyg~X-1 }
\def\cygp{Cyg~X-1}

\shorttitle{Gamma-ray emission from  Cygnus X-1} \shortauthors{Sabatini et al.}

\begin{document}

%% LaTeX will auto matically break titles if they run longer than

%\title{\bf Search and Detection of Transient Gamma-Ray Emission from the Microquasar Cygnus X-1}
\title{\bf Episodic Transient Gamma-Ray Emission from the Microquasar Cygnus X-1}


\bigskip

\author{S. Sabatini \altaffilmark{1,2},M. Tavani\altaffilmark{1,2}
E. Striani\altaffilmark{2},A. Bulgarelli\altaffilmark{5}, V.
Vittorini\altaffilmark{1,2},G. Piano\altaffilmark{1,2,11}, E. Del
Monte\altaffilmark{1}, M.~Feroci\altaffilmark{1},
F. de Pasquale\altaffilmark{20},
M.~Trifoglio\altaffilmark{5}, F.~Gianotti\altaffilmark{5},
 A.~Argan\altaffilmark{1},
G.~Barbiellini\altaffilmark{6},
% F.~Boffelli\altaffilmark{7},
P.~Caraveo\altaffilmark{3}, P.~W.~Cattaneo\altaffilmark{7},
A.~W.~Chen\altaffilmark{3,4},
% V.~Cocco\altaffilmark{1},
F.~D'Ammando\altaffilmark{1,2}, E.~Costa\altaffilmark{1}, G.~De
Paris\altaffilmark{1},G.~Di~Cocco\altaffilmark{5},
I.~Donnarumma\altaffilmark{1}, Y.~Evangelista\altaffilmark{1},
A.~Ferrari\altaffilmark{4,18},
 M.~Fiorini\altaffilmark{3},
% T.~Froysland\altaffilmark{2,4},
% M.~Frutti\altaffilmark{1},
F.~Fuschino\altaffilmark{5}, M.~Galli\altaffilmark{8},
A.~Giuliani\altaffilmark{3},M.~Giusti\altaffilmark{1},C.~Labanti\altaffilmark{5},
%I.~Lapshov\altaffilmark{1},
F.~Lazzarotto\altaffilmark{1},
P.~Lipari\altaffilmark{9}, F.~Longo\altaffilmark{6},
M.~Marisaldi\altaffilmark{5},
% M.~Mastropietro\altaffilmark{10},
S.~Mereghetti\altaffilmark{3}, E.~Morelli\altaffilmark{5},
E.~Moretti\altaffilmark{6}, A.~Morselli\altaffilmark{11},
L.~Pacciani\altaffilmark{1}, A.~Pellizzoni\altaffilmark{19},
F.~Perotti\altaffilmark{3}, P.~Picozza\altaffilmark{2,11},
M.~Pilia\altaffilmark{12,19},
%G.~Porrovecchio\altaffilmark{1},
G.~Pucella\altaffilmark{13}, M.~Prest\altaffilmark{12},
M.~Rapisarda\altaffilmark{13}, A.~Rappoldi\altaffilmark{7},
A.~Rubini\altaffilmark{1}, E.~Scalise\altaffilmark{1}, P.~Soffitta\altaffilmark{1},
 A.~Trois\altaffilmark{1},
E.~Vallazza\altaffilmark{6}, S.~Vercellone\altaffilmark{17},
A.~Zambra\altaffilmark{3}, D.~Zanello\altaffilmark{9},C.~Pittori\altaffilmark{14},
F.~Verrecchia\altaffilmark{14},
P.~Santolamazza\altaffilmark{14},P.~Giommi\altaffilmark{14}, S.~Colafrancesco\altaffilmark{14},
L.A.~Antonelli\altaffilmark{16}  L.~Salotti\altaffilmark{15}
}

\altaffiltext{1} {INAF/IASF-Roma, I-00133 Roma, Italy}
\altaffiltext{2} {Dip. di Fisica, Univ. Tor Vergata, I-00133 Roma,
Italy} \altaffiltext{3} {INAF/IASF-Milano, I-20133 Milano, Italy}
\altaffiltext{4} {CIFS-Torino, I-10133 Torino, Italy}
\altaffiltext{5} {INAF/IASF-Bologna, I-40129 Bologna, Italy}
\altaffiltext{6} {Dip. Fisica and INFN Trieste, I-34127 Trieste,
Italy} \altaffiltext{7} {INFN-Pavia, I-27100 Pavia, Italy}
\altaffiltext{8} {ENEA-Bologna, I-40129 Bologna, Italy}
\altaffiltext{9} {INFN-Roma La Sapienza, I-00185 Roma, Italy}
\altaffiltext{10} {CNR-IMIP, Roma, Italy} \altaffiltext{11} {INFN
Roma Tor Vergata, I-00133 Roma, Italy} \altaffiltext{12} {Dip. di
Fisica, Univ. Dell'Insubria, I-22100 Como, Italy}
\altaffiltext{13} {ENEA Frascati,  I-00044 Frascati (Roma), Italy}
\altaffiltext{14} {ASI Science Data Center, I-00044
Frascati(Roma), Italy} \altaffiltext{15} {Agenzia Spaziale
Italiana, I-00198 Roma, Italy} \altaffiltext{16}
{INAF-Osservatorio Astron. di Roma, Monte Porzio Catone, Italy}
\altaffiltext{17} {INAF-IASF-Palermo, via U. La Malfa 15, I-90146
Palermo, Italy}
 \altaffiltext{18} {Dip. Fisica, Universit\'a di Torino, Turin,
Italy} \altaffiltext{19} {INAF-Osservatorio Astronomico di
Cagliari, localita' Poggio dei Pini, strada 54, I-09012 Capoterra,
Italy}\altaffiltext{20} {ITAB, Via dei Vestini 33, I-66100 Chieti, Italy}

%\author{the AGILE Team, ASDC Team, E. Pian, R.F. Viotti, M.F. Corcoran,  .....(ORDER TBD)}
%\affil{}
%\and
%\author{R.F. Viotti,  L.A. Antonelli .....}
%\affil{} \and
%\author{E. Pian, ......}
%\affil{}


% \altaffiltext{1}{} \altaffiltext{2}{} \altaffiltext{3}{}
% \altaffiltext{4}{} \altaffiltext{5}{}


\begin{abstract}

Cygnus X-1 (\cygp) is the archetypal black hole (BH) binary
system in our Galaxy. We report the main results of
an extensive search for transient gamma-ray emission from Cygnus
X-1 carried out in the energy range 100 MeV -- 3 GeV by the {\it AGILE} 
satellite, during the period 2007 July -- 2009 October.
%AGILE observed the Cygnus region several times during the period 2007 July - 2009
%October, and accumulated an exposure time of about 300 days during
%which the source was mainly in the "hard" and "intermediate" X-ray
%spectral states.
The total exposure time is about 300 days, during which the source
was in the "hard" X-ray spectral state.
We divided the observing intervals in 2$\div$4 week
periods, and searched for transient and persistent emission.
We report an episode of significant transient
gamma-ray emission detected on 2009, October 16 in a position compatible with
\cyg optical position. This episode, occurred during a hard spectral
state of \cygp,
shows that a 1-2 day time variable emission above 100 MeV can be
produced during hard spectral states, having important theoretical implications
for current Comptonization models for \cyg and other microquasars.
Except for this one short timescale episode, no significant gamma-ray
emission was detected by {\it AGILE}. {\sas By integrating all available data we obtain a
2$\sigma$ upper limit for the total integrated flux of
$F_{\gamma,U.L.} =  3 \times 10^{-8} \rm \, ph \, cm^{-2} \,s^{-1}
$ in the energy range 100 MeV -- 3 GeV. We then clearly
establish the existence of a spectral cutoff in the energy range
1--100 MeV that applies to the typical hard state outside the flaring period
and that confirms the historically known spectral cutoff above 1 MeV.}

\end{abstract}


 %--- processes: black hole physics,
%acceleration of particle}

\setcounter{footnote}{1}

\section{Introduction}
\cyg is a binary system (discovered by Bowyer et al. 1965)
containing a O9.7 Iab supergiant star orbiting (5.6 days of period) % \cite{gies86})
around a compact star with a mass function of {\sas $ f= 0.23 \pm 0.01
\, M_{\odot}$ (Gies et al. 2008)} and a mass {\sas lower limit in the range 
$ 6\div 13$  $M_{\odot}$ ({\sass Zi{\'o}{\l}kowski} 2005)}.
%The compact star in Cyg~X-1 has a probable mass in excess of $10
%\, M_{\odot}$ when proper account is given to the orbital
%parameters and to the most probable companion star mass ($20-30 \,
%M_{\odot}$) \cite{gies86}.
\cyg is then the only known high-mass black hole (BH) binary system in our Galaxy (e.g.,
Tanaka \& Lewin 1995), and attracted considerable attention since
its initial mass range determinations \cite{bolton,webster}. Being
among the brightest X-ray binaries in our Galaxy (for a relatively
small distance of 2 kpc {\sas and average sub-Eddington X-ray luminosity
for a 10 solar mass compact object}), the system has
been extensively monitored in the radio, IR, UV and X-ray energy
bands (see Zdziarski \& {\sass Gierli{\'n}ski} 2004 for a review). 
%Theoretically, accretion
%processes onto a BH system are extensively studied and the highest
%detectable energies discussed using \cyg as a typical example
%(e.g., Zdziarski 1988; Gierlinski et al. 1999; Bednarek
%\& Giovannelli 2007; Zdziarski, Malzac \& Bednarek 2009).
%In particular, disk hydrodynamics and radiative
%and pair-creation properties of \cyg have been modelled with
%particular emphasis on the X-ray range and the highest detectable
%energies (e.g., Zdziarski 1988; Gierlinski et al. 1999; Bednarek
%\& Giovannelli 2007; Zdziarski, Malzac \& Bednarek 2009).

%Decades of \cyg high-energy observations\footnote{It is also
%important to note that \cyg is a (faint) radio source (of  few
%tens of mJy at 8 GHz) with interesting overall properties
%that correlate with the soft and hard X-ray emissions. Radio
%flares typically occur during state transitions or "failed state
%transitions" associated with intermediate states \cite{wilms}. }
%and monitoring by many instruments (including Ginga, OSSE-GRO,
%BSAX, RXTE, ASCA, XMM, Chandra, COMPTEL-GRO, INTEGRAL, Swift,
%Suzaku) firmly established the two main spectral ("hard" and
%"soft") states in the energy range between a few keV and a few MeV
%energy range.
 The system spends most of its time in the so
called "hard state" characterized by a
relatively low flux of soft X-ray photons (1--10 keV), a clear peak
of the photon energy spectrum in the hard X-ray band (around 100 keV), and
an energy cutoff around 1 MeV (e.g., {\sass Gierli{\'n}ski} et al. 1997,
McConnell et al. 2002, Del Monte et al., 2010). Occasionally, \cyg changes state
shifting its energy power spectrum to a "soft state" characterized
by a large flux in soft X-rays, a lower hard X-ray flux, and a tail
extending to energies up to 1 MeV and beyond \cite{mcconnell02}. 
\cyg is also detected in "intermediate
hard states", which usually show a less intense hard X-ray
emission and a shift of the spectral hump towards energies less
than 100 keV \cite{malzac06,wilms}. %In recent years Cyg X-1 was
%observed several times by INTEGRAL in these "intermediate states"
%that often, but not always, appear to occur when the source is about
%to switch from one state to the other.

Variability in \cyg above 100 keV was observed on several different time
scales, from months to {\sas milliseconds} (e.g. Brocksopp et al. 1999,
Ling et al. 1997, Pottschmidt et al. 2003, Zdziarski \& {\sass Gierli{\'n}ski} 
2004) and giant outburst episodes
have been detected in the 15--300 keV by the Interplanetary Network
\cite{Golenetskii03} during both spectral states.
%, maintaining the
% spectral parameters of the underlying state.

{\sas Theoretically, accretion processes onto a BH system are extensively studied 
using \cyg as a typical example. In particular, disk hydrodynamics and radiative
and pair-creation properties of \cyg have been modeled with
particular emphasis on the X-ray range and the highest detectable
energies (e.g., Zdziarski 1988; {\sass Gierli{\'n}ski} et al. 1999; Bednarek
\& Giovannelli 2007; Zdziarski, Malzac \& Bednarek 2009).
Extensive modelling of \cyg X-ray spectral
states have been carried out using Comptonization models (e.g.,
Titarchuk 1994; Poutanen \& Svensson 1996; Coppi 1999) and interpret
 the historical data available in the literature with a spectral cutoff 
near 1 MeV. 
%These
%models {\sas (mostly based on} inverse Compton emission by
%electrons (and pairs) energized in a hot corona with Thomson
%depth $\tau \sim 1$) usually
%predict a spectral cutoff near 1 MeV. 
Since the detection of a
non-thermal power law spectral component extending up to $\sim
1$MeV energies during the "soft" and "intermediate" states, the issue
of determining the variability and highest photon energies from
\cyg has been of crucial theoretical importance. A detection of
photon emission well above a few MeV from \cyg  would provide a
clear signature of efficient non-thermal acceleration processes
occurring in the system, that would need to be accounted for in
\cyg models and BH accretion disk modelling.}

Before the {\it AGILE} extensive monitoring of \cygp, only temporally
sparse information has been available in the energy range above a
few MeV. The gamma-ray instruments  on board of CGRO observed the
Cygnus region several times (typically with 2$\div$4 weeks long integrations)
during the period 1991-1997. In particular, the {\it EGRET} instrument
provided an overall upper limit to the flux of $ 10 \times 10^{-8}
\rm \, ph \, cm^{-2} \, s^{-1}$ above 100 MeV. {\it EGRET} observations
occurred always during "hard" spectral state, and did not cover at all
the "soft" state.

The only observation of {\it CGRO} during a soft state of \cyg was
carried out in June, 1996, following an X-ray alert provided by
{\it RXTE} \cite{cui97}. {\it OSSE} and {\it COMPTEL} observed \cyg from June 14 to
June 25, 1997 and this led for the first time to the detection of a
high energy power-law up to about 7 MeV
\cite{mcconnell02}. This indication of a power-law component extending to
 MeV and
beyond was also supported in recent
years by several {\it INTEGRAL} observations of \cyg \cite{cadolle06}.

A remarkable, although isolated, TeV flaring event of very
high-energy emission above $\sim$300 GeV from \cyg  was reported
by the {\it MAGIC} Cherenkov telescope during a set of observations in
2006 \cite{albert07}. The reported  VHE emission (for a pre-trial
significance above 4$\sigma$) was detected on 2006, September 24, for
about 1 hour (corresponding to an orbital
phase of 0.9) during a relatively bright hard X-ray emission
phase. Simultaneous {\it INTEGRAL} data \cite{malzac08} show that the TeV flare
from \cyg was detected $\sim1$day before an intense peak in hard
X-rays. However, at the time of the TeV flare, both the soft and
hard X-ray emission do not show significant variations or rapid
state changes: the spectral state was a "hard" one. This
detection of transient and very rapid TeV emission from \cyg
indicates that extreme particle acceleration processes may occur
also during a hard spectral state,
paving the way to detect non-thermal components also in states
previosly believed to be characterized by a cutoff above a few MeV.

In this \textit{Letter} we report
the {\it AGILE} search for short (days-weeks) timescale gamma-ray emission from \cyg
in the energy range 100 MeV -- 3 GeV with a total exposure time of $\sim300$ days,
during the period 2007 July -- 2009 mid-October.
Our data provide the first long timescale monitoring for
this important BH system. A separate paper (Del Monte et al.
2009) addresses the details of the X-ray emission as monitored by
{\it AGILE} and other detectors during our first year of observations.

\section{{\it AGILE} 2007-2009 Observations of Cygnus X-1 and data analysis}

%\begin{figure} %[t!]
%\begin{center}

%\includegraphics [width=7.5cm]{../B19b_CygX1_deep1145-12200FTint_100MeV_zoom.ps}

%\caption{\mt AGILE deep gamma-ray intensity map in Galactic coordinates
%of the Cyg X-1 region above 100 MeV for the data collected from
%2007 July to 2009 September. We overlayed the nominal position of Cyg X-1 with a
%white circle.
%The color bar scale is in units of $\rm photons \, cm^{-2}
%\, s^{-1} \, pixel^{-1}$. Pixel size radius
%is 0.1 degrees, and we used a 3-bin Gaussian smoothing. \\
%} \label{agiledeep_cygx1}
%\end{center}
%\end{figure}

The {\it AGILE} mission has been operating since 2007 April
 \citep{tavani-1}. The {\it AGILE} scientific instrument is very compact and is
characterized by two co-aligned imaging detectors operating in the
energy ranges 30 MeV -- 30 GeV ({\it GRID}, Barbiellini et al. 2002,
Prest et al. 2003) and 18--60 keV ({\it Super-AGILE}, Feroci et al.
2007), as well as by an anticoincidence system \citep{perotti} and
a calorimeter \citep{labanti}. {\it AGILE}'s performance is
characterized by large fields of view (2.5 and 1 sr for the
gamma-ray and hard X-ray bands, respectively) and optimal angular
resolution (PSF=$3^{\circ}$ at 100 MeV and PSF=$1.5^{\circ}$ at 400 MeV).
Flux sensitivity for a typical 1-week observing period can reach the
level of several tens of $ 10^{-8} \rm \, ph \, cm^{-2} \, s^{-1}
$ above 100 MeV, and 10--20 mCrab in the 18--60 keV range depending
on off-axis angles and {\mtav pointing directions} (see Tavani
et al. 2008 for
details about the mission and main instrument performance).

%%\begin{figure} %[t!]
%%\begin{center}

%%\includegraphics [width=8cm]{../12797-12810_FT_bin0.5_int_improved.ps}

%%\caption{\mt AGILE gamma-ray intensity map in Galactic coordinates
%%of the Cyg X-1 region above 100 MeV during the period
%%2009-10-15 UTC 23:13:36 to 2009-10-16 UTC 23:02:24 of the gamma ray flare.
%%The optical position of Cyg X1 is showed as a black circle and AGILE 2$\sigma$
%%confidence level contour in green. AGL J2022+4030 and AGL J2021+3652 sources,
%%used in the multi-source likelihood analysis of the field,
%%are also shown with black circles.
%%} \label{agile_cygx1}
%%\end{center}
%%\end{figure}

The {\it AGILE} satellite repeatedly pointed at the Cygnus region for a
total of $\sim$ 315 days ($\sim$ 13 Msec net exposure)
during the period 2007 July -- 2009 mid-October. The analysis of gamma-ray data presented in this
paper was carried out with the {\it AGILE-GRID} FT3ab$_{}$2$_{}$Build18
calibrated filter with a gamma-ray event selection that takes into
account South Atlantic Anomaly event cuts and 80$^{\circ}$ Earth
albedo filtering. Throughout the paper,
statistical significance assessment
and source flux determination was established using the
standard {\it AGILE} multi-source likelihood analysis software
\cite{chen10}. The method provides an assessment of
the statistical significance in terms of a Test Statistic (TS)
defined as in Mattox et al. 1996 and asymptotically distributed as a $\chi^2/2$ for
3 degrees of freedom ($\chi_3^2/2$).

\subsection{Search for persistent gamma-ray emission}

\begin{figure} %[t!]
\begin{center}
\includegraphics [width=9.cm]{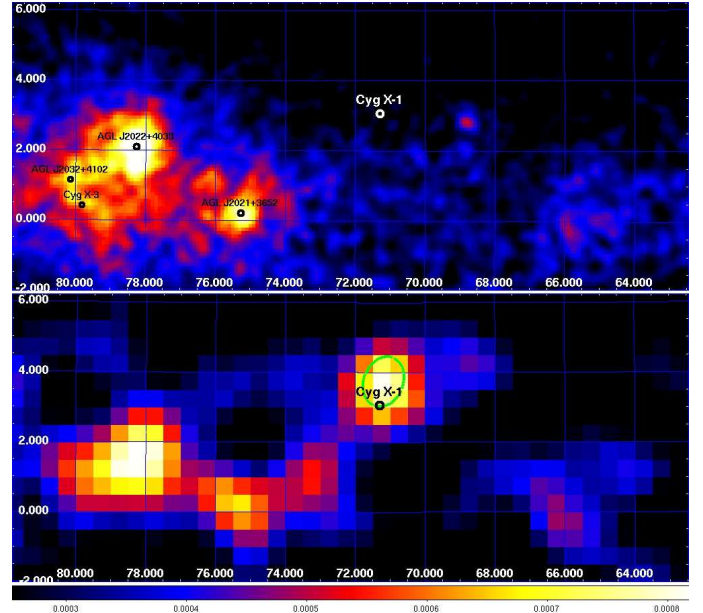}
\caption{\mt {\it AGILE} gamma-ray intensity maps above 100 MeV of the Cygnus Region
in Galactic coordinates displayed with a 3-bin Gaussian smoothing. \textit{Upper panel:} 
{\it AGILE} 2-years integrated map. Pixel size is
$0.1^{\circ}$. We overlayed the nominal position of Cyg
X-1 (white circle) and the other sources from {\it AGILE} catalogue.
The color bar scale is in units of $\rm
photons \, cm^{-2} \, s^{-1} \, pixel^{-1}$.
\textit{Lower panel:} {\it AGILE} 1-day map of the flaring episode of Cyg X-1
(2009-10-15 UTC 23:13:36 to 2009-10-16 UTC 23:02:24).
Pixel size is $0.5^{\circ}$. The black circle is the optical position of Cyg X-1
and the green contour is the {\it AGILE} 2$\sigma$ confidence level.
%AGL J2022+4030 and AGL J2021+3652 sources, used in the
%multi-source likelihood analysis of the field, are also shown with
%black circles.
} \label{agiledeep_cygx1}
\end{center}
\end{figure}

Multi-source likelihood analysis was used to search
for persistent emission from \cyg position in the
integrated sky map of the Cygnus region above 100 MeV for the
period 2007 July -- 2009 October (Fig \ref{agiledeep_cygx1}, upper panel).
The region is characterised by {\it AGILE} gamma-rays data 
showing two most prominent sources 1AGL
J2022+4032 and 1AGL J2021+3652 detected with high confidence
(38.8$\sigma$ and 24.6$\sigma$ respectively) and a gamma-ray flux
$F_{\gamma} = (123 \pm 4)  \times 10^{-8} \rm \, ph \, cm^{-2} \,s^{-1} $
and $F_{\gamma} = (57 \pm 4) \times 10^{-8} \rm \, ph \, cm^{-2}
\,s^{-1} $ respectively \cite{pittori09}. We also
detect Cygnus X-3 (3.2$\sigma$,
$F_{\gamma}= (10 \pm 3) \times 10^{-8} \rm \, ph \, cm^{-2} \,s^{-1} $;
Tavani et al, 2009), and the nearby
pulsar source 1AGL J2032+4102 (10.8$\sigma$, $F_{\gamma}= (35 \pm 3)
\times 10^{-8} \rm \, ph \, cm^{-2} \,s^{-1}$)\footnote{{\it AGILE} flux values 
are in agreement with the Fermi detections \cite{fermicat} for 
common sources.}. No statistically significant gamma ray source is
detected at a position consistent with that of Cyg X-1. The
2-sigma upper limit for the gamma-ray flux in the energy range 100 MeV -- 3 GeV is equal
to 3 $\times 10^{-8} \rm \, ph \, cm^{-2} \,s^{-1} $.

Data integrations of 2$\div$4 weeks exposure from single observation blocks
give typical 2-$\sigma$ upper limits in the range $(10-30) \times 10^{-8} \rm \, ph \,
cm^{-2} \,s^{-1}$.

\subsection{Search for transient gamma-ray emission}

Motivated by the X-ray variability of \cyg and by the particular
sequence of flaring gamma-ray emission from Cygnus X-3
\cite{tavnature}, we carried out a systematic search
for short (day) timescale variability of the \cyg  gamma-ray emission.
We used two independent and automatic methods for a blind search
of candidate gamma-ray transients in the region surrounding \cygp.

\vspace{0.2cm} 1.{\it The AGILE-GRID multi-source Likelihood
method.} The standard analysis pipeline uses a multiple source
likelihood analysis that iteratively optimizes position, flux and
significance of each source by successive repetitions in which the
parameters of one source are varied keeping all the others fixed.
This method is very efficient for relatively strong sources and
takes into account the Galactic diffuse emission and residual
background \cite{ab08}. It provides a pre-trial assessment of
statistical significance that needs to be corrected when used in
repeated systematic searches.
For this reason we also developed an independent method, that takes into
account multiple comparison corrections (see below).

\vspace{0.2cm} 2.{\it The False Discovery Rate Method (FDRM).} We
developed a detection method based on the False
Discovery Rate technique (FDR, Benjamini et al. 1995; Miller et
al. 2001, Hopkins et al. 2002) that is a statistical test
%\footnote{The data
%are tested against a given model, the null hypothesis,
% which in our case is the background count distribution.}
taking
into account the corrections for multiple testing, as
needed for example in repeated systematic searches. The FDRM allows
to control the expected rate of false detections (due to background
fluctuations) within a selected
sample. The method was adapted to the
analysis of {\it AGILE} gamma-ray data of the Galactic plane \cite{sabatini09}.
Given an observed distribution of background
counts-per-pixel (the null hypothesis), the selection is
based on choosing pixels characterized by
$p$-values\footnote{Given a statistical distribution, a
"$p$-value" assigned to a given value of a random variable is
defined as the probability, when the null hypothesis is true, of
obtaining that value or larger.} smaller than a threshold, $\alpha_{FDR}$.
%We
%define this set of pixels as the selected sample $S_p$ for a given
%$p$-value.
The crucial FDRM feature is that a $p$-value threshold
is \textit{not} fixed a priori (as in traditional statistical
methods), but is estimated on the data with the
requirement that
the rate of false detections, within the selected sample,
is the chosen $\alpha_{FDR}$ or smaller.
%It is customary to define this rate as the $\alpha$-FDR
%parameter
%\footnote{We note that $\alpha_{FDR}$ is not a confidence
%level, or a simple probability of occurrence. Rather, it is the
%ratio of the expected number of residual false detections that are
%to be found \textit{within} the selected sample $S_p$.}
%($\alpha_{FDR}$).
A typical value used in the literature \cite{miller01,hopkins}, and
that we adopt as a starting value for our search, is
$\alpha_{FDR}$=0.05. The FDRM ensures to control this rate, while
accounting for the 'post-trial' correction of a single detection
significance \cite{benjamini95}.
%In our case, multiple testing implies
%comparing the data (maps with a
%large number of pixels, or different observations of the same
%source) against a given model.
%The FDRM intrinsically takes into
%account the post-trial correction of a single detection
%significance \cite{benjamini95}.
%that allows to select candidate sources possibly present in
%gamma-ray counts maps of a given time integration.
We apply the FDRM in two different ways.

\vskip .3cm

\noindent \textit{The global-FDRM (G-FDRM)}: in this case we carry out a blind
search for (persistent or transient) sources in large
(\textit{global}) daily counts maps of the Galactic plane ($0.5^{\circ}$
pixel size). The null hypothesis for these daily maps is the
(background dominated) counts distribution of the Galactic plane
%(\textit{b}: $-5^{\circ}\div+5^{\circ}$ )
($\abs{b} \leq 5^{\circ}$). The random fluctuations and
the diffuse gamma-ray emission of these daily maps are well
described by  Poissonian distributions in {\it AGILE-GRID} data. Candidate
sources in the daily maps are identified as significant deviations
from the average distribution that applies to that specific day.
In our analysis we use a threshold of $\alpha_{FDR}=0.05$ that
limits the contamination by false positive sources in
the sample below 5$\%$.
%3x3 Pixel correlation due to GRID PSF
%at relevant energies was taken into account through a correction factor
%following Miller et al.2001.

\noindent \textit{The  source-FDRM (S-FDRM)}: the S-FDRM searches
for flaring episodes in the counts' light curve extracted from the
position of a \textit{single} candidate source location. In the
(verified) assumption that the average source flux at a given
position is typically below the
instrument sensitivity, unless it is producing (rare) flares,
the  null hypothesis in this case is obtained by
measuring the distribution of photon counts for the
specific sky location observed at intervals of 1-day. We considered
 the nominal \cyg position and used an aperture search radius of
1.5$^{\circ}$.
As in the case for G-FDRM, candidate flaring sources are detected as deviations
from the Poissonian average distribution, i.e. fluctuations with $p-$value below
the chosen threshold.
%\end{itemize}

\section{The gamma-ray flare of October 15-16, 2009}

All of the available {\it AGILE} data in the
archive from 2007 June to 2009 mid-October were searched for
variability on timescales of 1-day with both the Likelihood
and FDR methods. We used only data within 40$^{\circ}$
from the pointing direction and removed all data affected by
non-nominal satellite pointings.
%as assessed by specific multi-source likelihood analysis  is
%typically between $3\sigma$ to $4\sigma$. %, and below the
%significance threshold used in this paper.
%These candidate episodes do not pass our $5 \sigma$ threshold in order to qualify
%as a definite detection, and we do not report them in this
%\textit{Letter}. These episodes may constitute an interesting
%gamma-ray flaring activity related to intermediate spectral states
%of \cygp, and will be reported in detail elsewhere.
In this \textit{Letter} we consider only candidates
with at least 5$\sigma$ pre-trial significance.
Only one  gamma-ray flaring episode was definitely
detected in our thorough
search by both independent methods. The bottom panel of Fig~\ref{agiledeep_cygx1} shows the
{\it AGILE} gamma-ray intensity map above 100 MeV for this episode. The
emission  peaked during the time interval  2009 October, 15 (UTC
23:13:36) to 2009 October, 16 (UTC 23:02:24). The {\it AGILE-GRID}
multi-source likelihood analysis finds a TS=28.09 (=5.3$\sigma$
pre-trial,
4$\sigma$ post-trial\footnote{This corresponds
to a $p-$value of $1.7\times 10^{-6}$ (pre-trial) and $5.2\times 10^{-4}$ (post-trial).
% with 3 degrees of
%freedom. Applying the standard multiple comparison correction with n-trials=300 (days)
%we obtain a post-trial probability of $10^{-3}$ for this event to occur as a statistical
%fluctuation. For the $\chi^2$ distribution this corresponds to a statistical significance of $4\sigma$.
}, according to $\chi_3^2/2$ distribution and
multiple testing correction) detection
at the position (l,b)=71.2, 3.8 $\pm$ 0.7 (stat) $\pm$ 0.1 (syst)
consistent with the position of Cyg X-1, for a gamma-ray flux
of $F_{\gamma} = (232 \pm 66) \times 10^{-8} \rm \,
ph \, cm^{-2} \,s^{-1} $ in the energy range 100 MeV -- 3 GeV. The
detection is validated by both FDR methods: the G-FDRM analysis
finds the source with $\alpha_{FDR}=0.05$ and the S-FDRM analysis
with a highly significant $\alpha_{FDR}=0.001$. G-FDRM detection
has a lower significance due to the use of an average background
distribution which in this case overestimates the local background.
%These
%determinations correspond to pre-trial confidence levels of $6
%\times 10^{-5}$ and $2 \times 10^{-6}$, respectively.
%We carried out simulations of AGILE GRID daily counts maps and the occurrence
%rate of statistical fluctuations with the above given pre-trial
%confidence levels is less than 1 out of a 1000 days.
For comparison, during the same time
interval the source 1AGL J2022+4032 (Pittori et al. 2009), apparently coincident with
the SNR Gamma-Cygni, is detected with 3.1$\sigma$ significance with the likelihood analysis
(and $\alpha_{FDR-G}=0.05$), and a flux of $F_{\gamma} = (155 \pm
60) \times 10^{-8} \rm \, ph \, cm^{-2} \,s^{-1}$.  %It is interesting to note that this
The {\it Super-AGILE} (18--60 keV) flux
for \cyg for the day is  $F_{SA} = (580 \pm 48) $ mCrab and the ASM flux
is  $F_{ASM} = (268 \pm 20) $ mCrab in the 2--12 keV range.
The spectral state of the source was determined by means of the colour-colour
diagram obtained from ASM data as discussed in Del Monte et al. 2010.
Interestingly the flaring episode (MJD= 55120) occurred during a hard spectral state.
The orbital phase of \cyg was in the range 0.38-0.56. The system was detected to
subsequently evolve into one of the relatively rare dips of the
hard X-ray light curve.

%In several other occasions both automatic detection methods
%provided indications of episodic gamma-ray activity occurring at a
%position consistent with \cyg  during intermediate X-ray spectral
%states. However the statistical significance of these candidate
%episodes is just below our detection threshold and
%therefore we do not discuss them in this
%\textit{Letter}

%The gamma-ray flare detected by AGILE
%occurred during a transition towards a hard x-ray minimum.

%\begin{figure} %[t!]
%\begin{center}

%\includegraphics [height=5.5cm]{../BAT_CygX1_longterm+shortterm_lc_myplot_1.ps}
%\caption{\mt BAT long term daily light curve in the energy range 15-50 KeV.
%The black arrow shows the day of the gamma-ray flaring episode detected by AGILE.
%} \label{BATlc}
%\end{center}
%\end{figure}

\begin{figure} %[t!]
\begin{center}

% \includegraphics [height=12.5cm]{fig1.eps}
% \includegraphics [height=12.5cm]{12797-12810_FT_bin0.5_int.jpg}

%\vspace{8cm}%
%\includegraphics [width=8.3cm]{ASM_BAT_longterm_lc_v2.eps}
\includegraphics [width=9.cm]{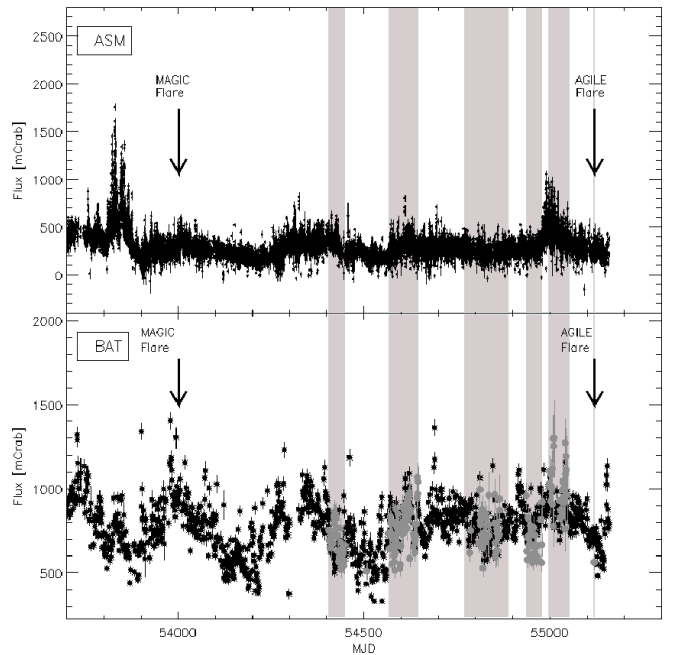}

\caption{\mt \textit{Upper panel}: {\it RXTE/ASM} daily light curve of
\cyg over the period 2005 November to 2009 October, in the energy
range 2--10 keV. The black arrows show the day of {\it MAGIC} and {\it AGILE} flaring
episode. Gray regions are {\it AGILE} pointings of the Cygnus Region.
 \textit{Lower panel}: {\it Swift BAT} long term daily light
curve in the energy range 15--50 keV and {\it Super-AGILE} data (gray dots) when available.  }
\label{BATlc}
\end{center}
\end{figure}

%\begin{figure} %[t!]
%\begin{center}

% \includegraphics [height=12.5cm]{fig1.eps}
% \includegraphics [height=12.5cm]{12797-12810_FT_bin0.5_int.jpg}
%\includegraphics [height=5cm]{../mcconnell2002_spettrocygx1.ps}

%\caption{\mt McConnell et al. 2002. Cyg X-1 spectral states.
%} \label{ASMlc}
%\end{center}
%\end{figure}

\begin{figure*} %[t!]
\begin{center}

\includegraphics[width=9.5cm]{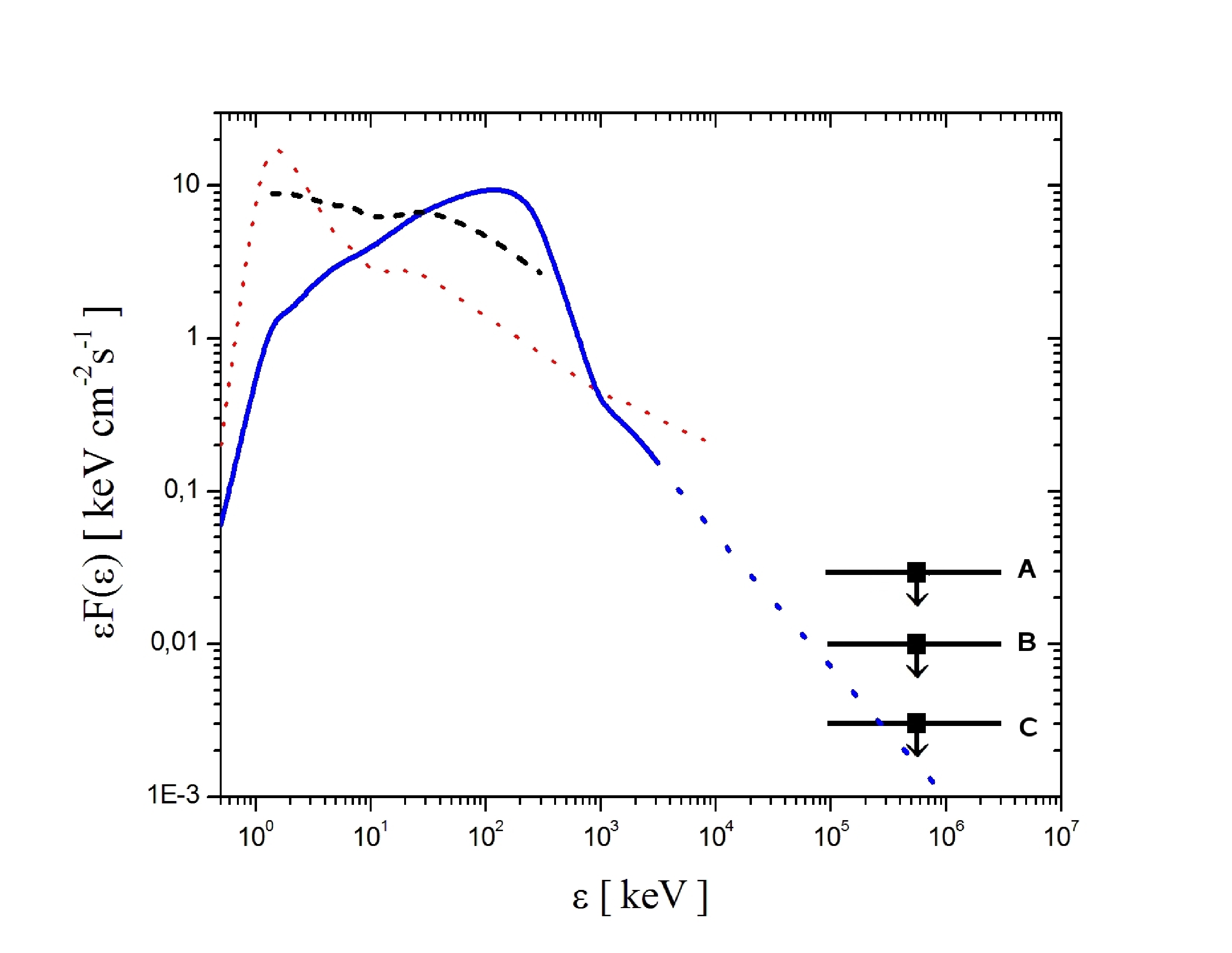}\\
\includegraphics[width=9.5cm]{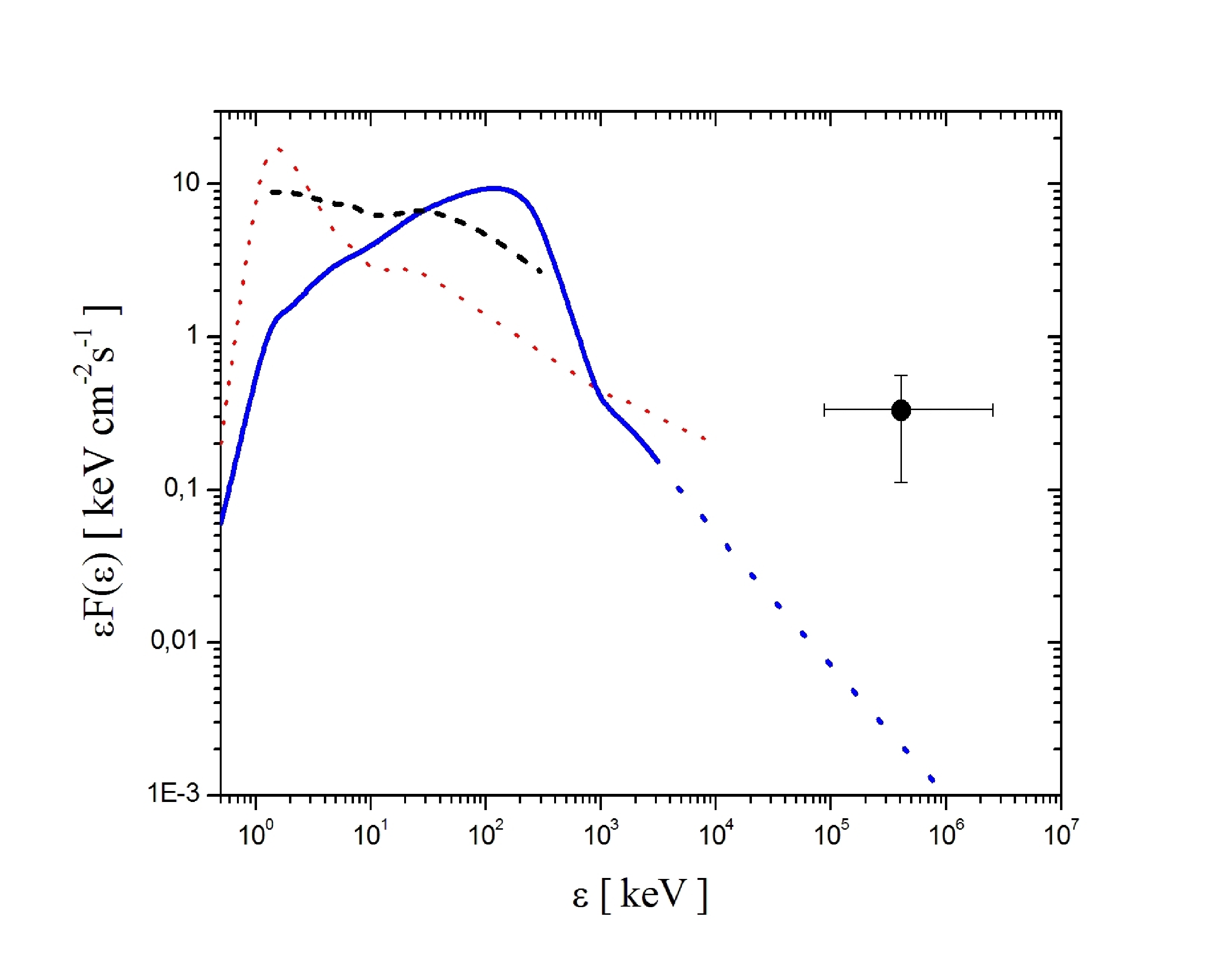}

\caption{\mt Cyg X-1 spectral energy distribution in typical states
(``hard'' in solid line, ``soft'' in dotted line, ``intermediate'' in dashed line;
%,  with dotted line extrapolation to higher energies
{\sass Gierli{\'n}ski} et al. 1999, Zdziarski et al. 2002). The dashed line extrapolated from the
hard X-ray state is a purely graphical extension of the trend suggested by the
historical data.
\textit{Upper panel}: {\it AGILE} 2$\sigma$ upper limits above 100 MeV
for integration times of: 2 weeks (A), 4 weeks (B) and $\sim 315$ days (C). \textit{Lower panel}: 
{\it AGILE} data above 100 MeV for the flaring episode.
%  We take typical spectral states from Zdziarski et al. 2002.
  %the RXTE spectrum of 1996 May 23 (reported by Gierlinski et al. 1999;
%Zdziarski et al. 2002)
%as a typical intermediate state spectrum.
}
\label{vFv}
\end{center}
\end{figure*}

\section{Discussion}

Fig \ref{BATlc} shows the X-ray historical light curves of \cyg from 2005 
November: the upper panel reports the {\it RXTE/ASM} data,  and
the lower panel the {\it Swift/BAT} data,
superimposed with the {\it Super-AGILE}
data from 2007 November (gray dots).
Gray zones highlight {\it AGILE} pointings of the
Cygnus region. The {\it ASM} data show that after MJD~53900 the system
did not undergo clear transitions to one of its soft states anymore. The {\it Swift/BAT}
hard X-ray data are available for the last 4 years and show a
pattern with rare dips occurring almost once a year. 
%It is
%interesting to note that the {\it MAGIC} TeV flare was detected during a
%hard state.

The {\it AGILE} dataset extends for $\sim$300 days, during which the system
was in its typical hard X-ray state ~\cite{delmonte}.
The lack of relatively strong gamma-ray
emission on a timescale of weeks together with the deep upper
limit obtained by integrating all {\it AGILE-GRID} data clearly confirms
the existence of a spectral cutoff between 1 and 100 MeV in the
typical hard state. Fig.~\ref{vFv} (upper panel)
shows the spectral energy distribution of \cyg  with its
typical historical spectral states
%(hard in solid curve,
%soft in dotted curve, intermediate in dashed curve).
In the same figure,
typical {\it AGILE} upper limits are given for 2-, 4-weeks and $\sim 300$ days
integrations. This gamma-ray average spectral behaviour of \cyg in
the hard state during week-month timescales is in
overall agreement with Comptonization models {\sas of black hole candidates (e.g.,
Titarchuk 1994; Poutanen \& Svensson 1996; Coppi 1999) and more specifically of Cyg X-1
({\sass Gierli{\'n}ski} et al. 1997; McConnell et al. 2002).}

%  during a typical pointing of about 30 days($ 10
% \times 10^{-8} \rm \, ph \, cm^{-2} \, s^{-1}$ above 100 MeV)
% clearly establishes that both spectral states must have a cutoff
% at energies above 10 MeV.

However, %if
our detection of October 16, 2009 %is associated
%with \cyg, this would be
is the first reported 1-day
gamma-ray flare in the energy range 100 MeV -- 3 GeV from
the system during a
hard state. This shows that
physical processes can occasionally be more complex than predicted by
current models. The
lower panel of Fig.~\ref{vFv} shows the {\it AGILE-GRID}
gamma-ray detection during such flare together with the
spectral shapes characterizing the different spectral states
for reference.
Efficient particle acceleration
occurs also in states characterized by the
presence of a hot corona that should be in pair-Comptonized
equilibrium (e.g., Zdziarski 1988; Zdziarski et al. 2009). The
gamma-ray emission can have leptonic or hadronic origin (e.g.,
Perucho \& Bosch-Ramon 2008), depending on the model as well as on the
assumptions on the acceleration site (close or far from the
inner disk and/or jet). Lack of simultaneous TeV data prevents a
more complete spectral analysis of the gamma-ray flaring event. We
note that the TeV spectrum reported by {\it MAGIC} occurred also during a hard 
state and having a photon spectral index $\alpha = 3.2 \pm 0.6$ \cite{albert07} is in
qualitative agreement with our {\it AGILE} spectral detection\footnote{{\sas
The gamnma-ray detection has a flux about a factor three above 
the model of Zdziarski et al. 2009.}}, even
though the broad-band spectrum may be complex and have several
independent components. A theoretical analysis of our results is
well beyond the scope of this paper.

\section{Conclusions}

{\it AGILE} extensive monitoring of \cyg in the energy range 100 MeV -- 3 GeV
during the period 2007 July -- 2009 October confirmed the
existence of a spectral cutoff between 1--100 MeV during
the typical hard spectral state of the
source. However, even in this state, \cyg is
capable of producing episodes of extreme particle acceleration on
1-day timescales. Our first
detection of a gamma-ray flare above 100 MeV adds to the
even shorter detection in the TeV range by {\it MAGIC}. These data have
great relevance for a more detailed theoretical modeling of pair
equilibrium Comptonized coronae and non-thermal particle
acceleration that may co-exist for short timescales of order
of hours-days.

We note that the gamma-ray flaring activity detected by {\it AGILE} from
\cyg during its decreasing trend of hard X-ray emission is
qualitatively similar (transition to a hard x-ray minimum)
to what observed in the case of the other
microquasar Cygnus X-3 \cite{tavnature,fermiscience}. Whether this behavior is
common to microquasars and BH accreting systems is a fascinating
question that will be addressed by future observations.

\section{Acknowledgements}
We thank the anonymous referee for the contribution to the improvement 
of our manuscript. The {\it AGILE} mission is funded by the Italian Space Agency with scientific
and programmatic participation by the Italian Institute of Astrophysics
and the Italian Institute of Nuclear Physics.


\begin{thebibliography}{}

\bibitem[Abdo et al. 2009a]{fermicat}
Abdo A.A. et al., 2009, ApJS,183, 46A

\bibitem[Abdo et al. 2009b]{fermiscience}
Abdo A.A. et al., 2009, Science, 326, 1512

\bibitem[Albert et al. 2007] {albert07}
Albert J. et al, 2007, ApJ, 665, L51

\bibitem[Barbiellini et al. 2002] {barbiellini02}
Barbiellini G. et al, 2002, NIM A, 490, 146

\bibitem[Bednarek \& Giovannelli 2007]{bednarek}
Bednarek, W. \& Giovannelli, F., 2007, A\&A, 464, 437

\bibitem[Belloni 2009]{belloni}
Belloni, T.M., 2009, to appear in Belloni, T. (ed.):
    \textit{The Jet Paradigm - From Microquasars to Quasars}, Lecture Notes Phys. 794; arXiv:0909.2474

\bibitem[Benjamini \& Hochberg 1995]{benjamini95}
Benjamini Y., \& Hochberg, Y., 1995, J. R. Stat. Soc. B, 57, 289

\bibitem[Bolton 1972]{bolton}
Bolton, C.T., 1972, Nature, 235, 271

\bibitem[Bowyer et al., 1965]{bowyer65}
Bowyer, S.C., Byram, E. T., Chubb, T. A., Friedman, M.,  1965, Science, 147, 394

\bibitem[Brocksopp et al. 1999]{brocksopp99}
Brocksopp C. et al, 1999, MNRAS, 309, 1063

\bibitem[Bulgarelli et al. 2008]{ab08}
Bulgarelli et al, 2008, in \textit{Astronomical Data Analysis Software and Systems XVIII},
ASP Conference Series, Vol XXX

\bibitem[Cadolle Bel et al. 2006]{cadolle06}
Cadolle Bel et al, 2006, A\&A, 446, 591

\bibitem[Chen et al. 2010]{chen10}
Chen et al, 2010, in preparation

\bibitem[Coppi 1999]{coppi}
Coppi, P.S., 1999, in \textit{High Energy Processes in Accreting
Black Holes}, ASP Conference Series 161, ed. J. Poutanen \& R.
Svensson, p.375

\bibitem[Cui et al. 1997]{cui97}
Cui et al., 1997, ApJ, 474, L57

\bibitem[Del Monte et al. 2010]{delmonte}
Del Monte E., et al., 2010,A\&A, submitted

\bibitem[Feroci et al. 2007]{feroci}
Feroci M., et al., 2007, NIM A, 581, 728

\bibitem[Gierli{\'n}ski et al. 1997]{gierlinski97}
Gierli{\'n}ski, M. et al., 1997, MNRAS, 288, 958

\bibitem[Gierli{\'n}ski et al. 1999]{gierlinski}
Gierli{\'n}ski, M., Zdziarski, A.A., Poutanen, J., Coppi, P.S.,
Ebisawa, K. \& Johnson, W.N., 1999, MNRAS, 309, 496

\bibitem[Gies \& Bolton 1986]{gies86}
Gies D.R. \& Bolton C.T., 1986, ApJ, 304, 371

\bibitem[Gies et al. 2008]{gies08}
Gies D.R. et al., 2008, ApJ, 678, 1237

\bibitem[Golenetskii et al. 2003]{Golenetskii03}
Golenetskii S. et al, 2003, ApJ, 596, 1113

\bibitem[Hopkins et al. 2002]{hopkins}
Hopkins, A.M., et al., 2002, AJ, 123, 1086

\bibitem[Labanti et al. 2006]{labanti}
Labanti C. et al., 2006, proc SPIE, 6266, 62663

\bibitem[Ling et al. 1997]{ling97}
Ling J.C., et al., 1997, ApJ, 484, 375

\bibitem[Malzac et al. 2006] {malzac06}
Malzac J. et al, 2006, A\&A, 448, 1125

\bibitem[Malzac et al. 2008] {malzac08}
Malzac J. et al, 2008, A\&A, 492, 527

\bibitem[McConnell et al. 2000]{mcconnell00}
McConnell M.L. et al., 2000, ApJ, 543, 928

\bibitem[McConnell et al. 2002]{mcconnell02}
McConnell M.L. et al., 2002, ApJ, 572, 984

\bibitem[Miller et al. 2001]{miller01}
Miller C.J. et al., 2001, ApJ, 122, 349

\bibitem[Perotti et al. 2006]{perotti}
Perotti F. et al, 2006, MIN A, 556, 228

\bibitem[Perucho \& Bosh Ramon 2008]{perucho08}
Perucho M. \& Bosh-Ramon V., 2008, A\&A, 482, 917

\bibitem[Pittori et al. 2009]{pittori09}
Pittori, C., e al., 2009, A\&A, 506, 1563

\bibitem[Poutanen \& Svensson 1996]{poutanen}
Poutanen, J. \& Svensson, R., 1996, ApJ, 470, 249

\bibitem[Pottschmidt et al. 2003]{pottschmidt03}
Pottschmidt et al, 2003, A\&A, 407, 1039

\bibitem[Prest et al. 2003]{prest03}
Prest M. et al. 2003, NIM A, 501, 280

\bibitem[Sabatini et al. 2010]{sabatini09}
Sabatini et al, 2010, in preparation

\bibitem[Tanaka \& Lewin 1995]{tanaka}
Tanaka Y. \& Lewin, W.H.G., 1995, in  \textit{X-Ray Binaries},
edited by W.H.G. Lewin, J. van Paradijs, E.P.J. van den Heuvel
(Cambridge, Cambridge University Press), p. 126

\bibitem[Tavani et al. 2008]{tavani-1}
Tavani, M., et al. 2008, A\&A, 502, 995

\bibitem[Tavani et al. 2009] {tavnature}
Tavani, M., et al. 2009, Nature, 462, 620

\bibitem[Titarchuk 1994]{titarchuk}
Titarchuk, L., 1994, ApJ, 434, 570

\bibitem[Webster \& Murdin 1972]{webster}
Webster, B.L., \& Murdin, P., 1972, Nature, 235, 37

\bibitem[Wilms et al. 2006]{wilms}
Wilms, J., Nowak, M. A., Pottschmidt, K., Pooley, G. G., Fritz,
S., 2006, A\&A, 447, 245

\bibitem[Zdziarski 1988]{zdziarski88}
Zdziarski, A~.A., 1988, ApJ, 335, 786

\bibitem[Zdziarski et al. 2002]{zdziarski02}
Zdziarski, A.~A., Poutanen J., Paciesas W.~S., Wen L., 2002, ApJ, 578, 357

\bibitem[Zdziarski \& Gierli{\'n}ski 2004]{zdziarski04}
Zdziarski, A.~A. \& Gierli{\'n}ski, M., 2004 Prog. Theor. Phys. Suppl., No. 155, 99

\bibitem[Zdziarski, Malzac \& Bednarek 2009]{zdziarski09}
Zdziarski, A~.A., Malzac, J., \& Bednarek, W., 2009, MNRAS, 394,
L41

\bibitem[Zi{\'o}{\l}kowski 2005]{ziolkowski05}
Zi{\'o}{\l}kowski J., 2005, MNRAS, 358, 851

\end{thebibliography}
\end{document}